\documentstyle[12pt,dina4,preprint,psfig,aps]{revtex}
\begin{document}
\draft
\title{ Neural Networks for Impact Parameter Determination
\footnote{supported by GSI,BMBF and DFG} }

\author{
        S. A. Bass, A. Bischoff, J. A. Maruhn,
	H. St\"ocker and W. Greiner}
\address{
  	Institut f\"ur Theoretische Physik \\
	Johann Wolfgang Goethe Universit\"at\\
	Robert Mayer Str. 8-10\\
	D-60054 Frankfurt am Main, Germany
}

\maketitle

\begin{abstract}
An accurate impact parameter determination in a heavy ion collision is crucial
for almost all further analysis. The capabilities of an
artificial neural network are investigated to that respect.
A novel input generation for the network is proposed, namely
the transverse and longitudinal momentum distribution
of all outgoing (or actually detectable) particles.
The neural network approach
yields an improvement in performance of a factor of two as compared to
classical techniques.
To achieve this improvement simple network architectures and
a $5\times 5$ input grid in ($p_t,p_z$) space are sufficient.
\end{abstract}

\pagebreak

The physics of relativistic heavy ion collisions is motivated
by the unique opportunity to study
the properties of hot and dense nuclear
matter \cite{sch68,sch74,mar85,cse86,sto86,st86,cl86,schue87,cas90}.

For a detailed investigation of heavy ion collisions a proper event
characterization is mandatory. In particular, the impact parameter, though
not directly accessible experimentally, is among the most
important characteristics for the description of the event geometry
and selection.

For the investigation of highly compressed nuclear matter,
it is important to select the most
central collisions. On the other hand, recently discovered new phenomena
such as pionic bounce--off and squeeze--out are only observed in
semi-peripheral collisions.

There have been various proposals how to determine the impact
parameter. Most of these are based on the mean particle multiplicity,
the ratio of transverse to longitudinal energy deposition,
a transverse momentum analysis (directivity-cut) or on a combination
of these techniques.
However, all previously mentioned methods have one thing in common:
They tend to break down for very central collisions and are generally
optimized for a certain impact parameter range. The achievable
accuracy is estimated to be at best $\pm 1$ to $\pm 1.5$ fm.

Recently,  neural networks have been suggested as tools for impact parameter
determination \cite{ba94n,ai95n}. While offering an improvement of
approximately 50 \% as compared with ``classical'' techniques, their main
shortcoming was the large amount of preprocessing neccessary
for the selected input.
Essentially, the proposed networks were used as multidimensional nonlinear
fits to combine the (already known) impact-parameter dependence
of the various input observables (a combination of
three observables has been used) into a single function yielding the
impact parameter.

The use of preprocessed input has the advantage of
reducing the amount of input data
(and therefore the computing time for the network output).
It's great disadvantage, however, is that only those correlations and
informations preselected for the input can be accessed and
optimized by the network.
Previously unknown correlations between input observables and the
desired output may be destroyed or left out through the preprocessing.
Minimizing the amount of preprocessing
allows for taking advantage of the full capabilities of the neural network.
In addition, common preprocessing techniques are computationally expensive.

Those heavy ion collision events which are used as input for
training and analyzing the
network's performance have to be supplied by a
theoretical model
rather than by experiment (otherwise it would be impossible
to compare the network output with a target value for the
impact parameter).

For the present study
an extension of the {\bf Q}uantum {\bf M}olecular
{\bf D}ynamics model (QMD) \cite{ai86,ai87b,ai91,ba95} has been applied.
It explicitly
incorporates isospin and pion production via the delta resonance (IQMD)
\cite{ha88,ha89,hadiss}.
In the QMD model the nucleons are represented by Gaussian
shaped density distributions, giving the nuclei a surface thickness of
1.5 fm.
The initial momenta are randomly chosen between 0 and the local
Thomas-Fermi-momentum.
The $A_P$ projectile and $A_T$ target nucleons interact via two- and
three-body
Skyrme forces, a Yukawa potential, momentum dependent interactions, a
symmetry potential (to achieve a correct distribution of protons and neutrons
in the nucleus) and explicit Coulomb forces
between the $Z_P$ and $Z_T$ protons.
They are propagated according to Hamilton's equations of motion.
Hard N-N-collisions are included by employing
the collision term of the well known VUU/BUU
model \cite{st86,kru85a,ai85a,wo90,lib91a}. The collisions are done
stochastically, in a similar way as in the cascade models \cite{yar79,cug80}.
In addition, the Pauli blocking (for the final state) is taken into account
by regarding the phase space densities in the final states of a two-body
collision. Clusters (e.g. deuterons and tritons) are formed via
a configuration space coalescence model.

\begin{figure}
\centerline{\psfig{figure=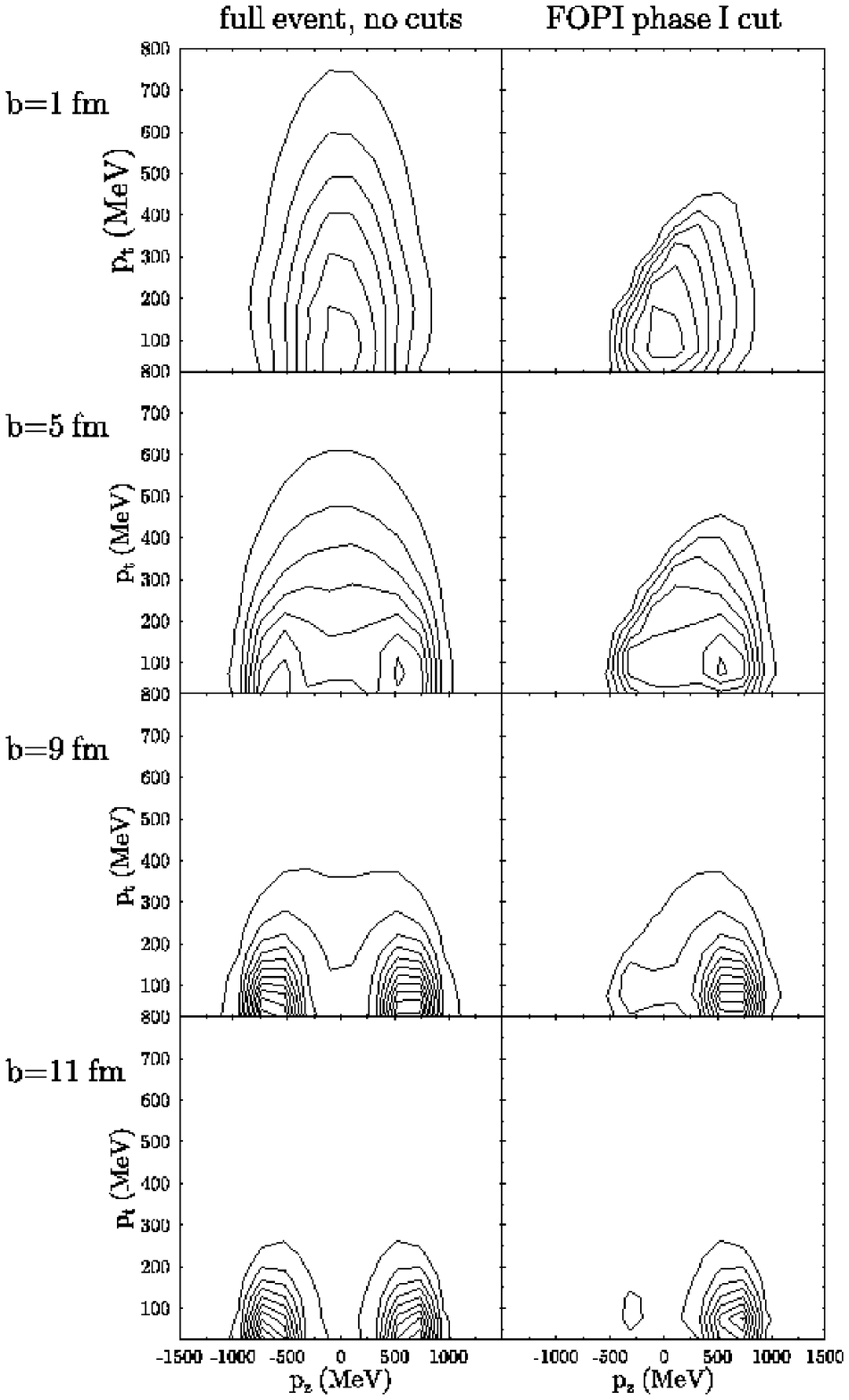,height=15cm}}
\caption{\label{pspace} $\frac{dN}{p_t dp_t dp_z}$ (all momenta per nucleon)
for nucleons and fragments in  Au+Au collisions at 1 GeV/nucleon
and in the final state, i.e. after all collisions have ceased.
The left column shows the full $4 \pi$ acceptance whereas the right
column shows the events filtered with the acceptance of the FOPI detector
(phase I). The different rows depict different impact parameter bins.}
\end{figure}

In order to minimize preprocessing and to allow the network to use
``unknown correlations'' the longitudinal
and transverse momenta per nucleon of all baryons in the system were
choosen as input.
The left column of figure \ref{pspace} shows the respective final state
momentum distributions for impact parameters of 1, 5, 9 and 11 fm.
In contrast to the full $4\pi$ information accessible via a QMD simulation,
experiments usually have a limited acceptance. The right column of
figure \ref{pspace} shows the same momentum distributions as the left
column, this time processed with the detector filter \cite{fopifilt}
of the FOPI experiment \cite{fopirefs} at GSI.  The filter does not only
destroy the symmetry to the $p_z=0$ axis, but it also reduces drastically
the available phase space information.
The suggested neural algorithms should prove their usefulness both for
the ``ideal'' $4\pi$ dataset and for the more ``realistic''
filtered set. A simplistic extrapolation of the network performance (from the
``ideal'' set to the ``realistic'' one), as performed
for the preprocessed input of IMF-multipicity, $p_{x,dir}$ and ERAT
in \cite{ai95n}, is clearly not justified.

The event data have been transformed
into a constant number of input components to be used with neural network.
In this paper a new direct mapping of the momentum-state
distribution to the network's input is proposed.
The momentum distributions is discretized in a two-dimensional grid.
For each of the ($3\times 3$ to $20\times 20$, equally sized)
momentum bins, the corresponding number of ``hits'' is calculated
and used as one input component.

We now sketch the neural network algorithm used,
a standard feed-forward two-layer perceptron trained by
error-backpropagation \cite{rum,mu90}.
The network
consists of a ``hidden'' layer of up to 20 nonlinear units receiving inputs
from the applied data vector and transferring
their signals to the output unit (a single linear output unit is used,
whose continuous valued output represents the impact parameter).

Every (hidden or output) unit performs a weighted
sum over all input signals. The hidden units calculate their own signals
by applying a nonlinear ``squashing'' function $\sigma(x)$ to the result:
\begin{displaymath}
y_j^{\rm hid} = \sigma(x_j^{\rm hid})
= \sigma\left(\sum_k w_{jk}^{\rm hid}\,y_k^{\rm in}\right)\quad.
\end{displaymath}
As a squashing function $\sigma(x) = \tanh(x)$ is used.
$y_k^{\rm in}$ is a component of the data vector.
A connection weight (from input component $k$ to hidden unit $j$)
is given by $w_{jk}^{\rm hid}$.
The output unit is linear:
\begin{displaymath}
y^{\rm out} = \sum_j w_j^{\rm out}\,y_j^{\rm hid}\quad.
\end{displaymath}
For each unit, a connection to a constant signal, $y_0\equiv 1$, is included
which provides an activity threshold.

First, the network's weights are initialized with small random values.
For each learning pattern an output is produced during training. It is  rated
by the error function
\begin{displaymath}
E = \frac{\displaystyle 1}{\displaystyle 2}\left(\delta^{\rm out}\right)^2\; ,
\end{displaymath}
where
\begin{displaymath}
\delta^{\rm out} = y^{\rm target}-y^{\rm out}
\end{displaymath}
with $y^{\rm target}$ as the desired output. Successively for each pattern,
the weights are updated according to a gradient descent in the weight space
with respect to the error function,
\begin{displaymath}
\Delta w_l^{\rm out} \; = \; -\varepsilon\,\frac{\displaystyle \partial E}
	{\displaystyle \partial w_l^{\rm out}}\; ,\qquad
\Delta w_{lm}^{\rm hid} \; = \; -\varepsilon\,\frac{\displaystyle \partial E}
	{\displaystyle \partial w_{lm}^{\rm hid}}\; ,
\end{displaymath}
with $\varepsilon > 0$ as learning rate. This leads to the learning rules
\begin{displaymath}
\Delta w_l^{\rm out}
 = \varepsilon\,\delta^{\rm out}\,y_l^{\rm hid}
\end{displaymath}
and
\begin{eqnarray*}
\Delta w_{lm}^{\rm h}
& = & -\varepsilon \; \frac{\displaystyle \partial}
	{\displaystyle \partial w_{lm}^{\rm hid}}
\left[\frac{\displaystyle 1}{\displaystyle 2}\left( y^{\rm target}
- \sum_j w_j^{\rm out}\,y_j^{\rm hid}\right)^2\right]\nonumber\\
	{\displaystyle \partial w_{lm}^{\rm h}}\nonumber\\
& = & \varepsilon\,\delta^{\rm out}\,
w_l^{\rm out}\,\sigma'(x_l^{\rm hid})\,y_m^{\rm in}\nonumber\\
& = & \varepsilon\,\delta_l^{\rm hid}\,y_m^{\rm in}\; ,
\end{eqnarray*}
where
\begin{displaymath}
\delta_l^{\rm hid}
 = \delta^{\rm out}\,w_l^{\rm out}\,\sigma'(x_l^{\rm hid})\; .
\end{displaymath}
The weight update values $\Delta w$ are added to the corresponding weights
directly after each presentation of a learning sample
(incremental learning).

The event data generated by the theoretical model were divided into three
sets of equal size.
The first set was used as training data, the second provided a
criterion for stopping the learning process (in order to avoid overfitting),
whereas the third one was used to determine the
generalization ability of the system.
Since there were virtually no limitations to the number of events available,
large sets (1000 events each) could be used to reduce fluctuation of
performance.
For a complete learning session, typically several hundred cycles through
the entire training data were necessary.
Training was stopped when the performance on the criterion data set
was best.

In contrast to the extensive training phase,
very little calculation time is needed for the application
of a trained network, so that it could be integrated into the data analysis
process itself.

\begin{figure}
\centerline{\psfig{figure=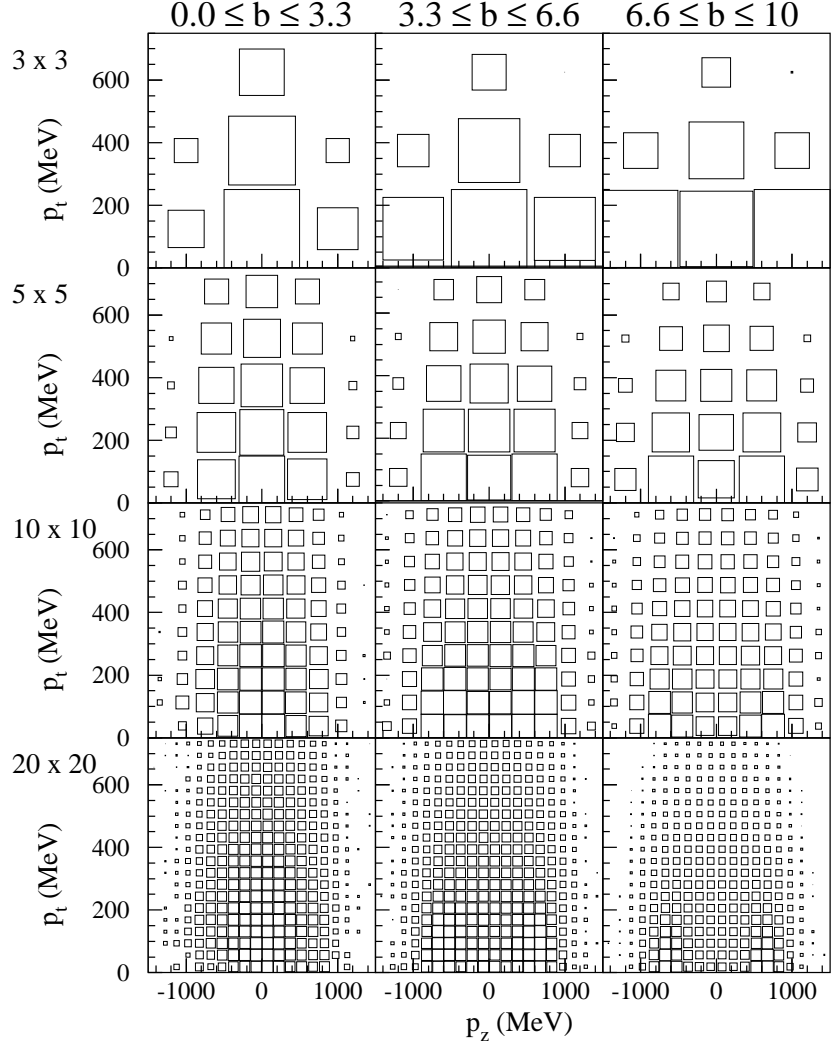,height=15cm}}
\caption{\label{pattern} Discretization of $\frac{dN}{p_t dp_t dp_z}$
for input into the neural network. The columns refer to central,
medium and peripheral collisions (left to right). The rows show
the input for different grid dimensions (top to bottom: 3, 5, 10 and 20).
The area of the squares is proportional to the number of particles in
the bins.
 }
\end{figure}

\begin{figure}
\centerline{\psfig{figure=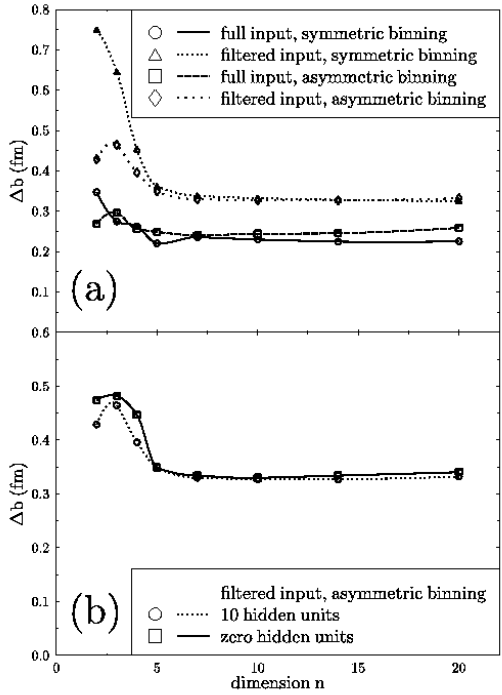,height=15cm}}
\caption{\label{perform} Network performance  $\Delta b$ versus
input grid dimension $n$. The upper frame compares calculations with and
without detector filter.
The original boundaries which cover the available phase space information
in the unfiltered case are marked as {\em symmetric binning} whereas
the boundaries adapted to the detector acceptance of the filtered data-set
are marked as {\em asymmetric binning}.
In the lower frame calculations
with and without hidden units are displayed.}
\end{figure}

Figure \ref{pattern} shows the effects of the chosen resolution (the number
of input components) on the selected input for three impact parameter bins
(central, medium and peripheral impact parameters).
When increasing the resolution in order to improve the informational contents
of the input the following points have to be taken into account:
\begin{itemize}
\item The computational time rises quadratically
with the input grid dimension and therefore
the processing time poses a constraint for the resolution.
\item The increase in information through an increased resolution
does not necessarily translate into an improved network
performance (see figure \ref{perform}).
\end{itemize}

\begin{figure}
\centerline{\psfig{figure=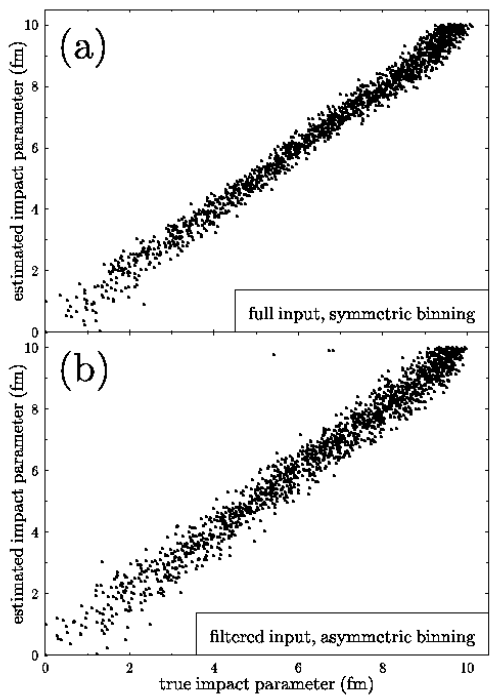,height=15cm}}
\caption{\label{result1} True impact parameter versus the reconstructed
impact parameter for full events (upper frame) and events processed
with the filter of the FOPI experiment  (lower frame).
Each dot represents one event. }
\end{figure}

Sample results can be seen in figure \ref{result1}.
The true impact parameter is plotted versus the reconstructed
impact parameter for full events (upper frame) and events processed
with the filter of the FOPI experiment\cite{fopifilt} (lower frame).
Each dot represents one event.
Both calculations have been performed with a $10 \times 10$ input grid,
10 hidden units, and one output unit. In the lower frame the boundaries
in $p_t$ and $p_z$ have been adapted to the acceptance range of the
detector filter. In both cases the quality of reconstruction remains
constant for the
whole impact parameter range.

The effect of the detector filter can
be seen in the slight broadening of the distribution.
Although the symmetry along the $p_z=0$ axis has been broken and
phase space severely reduced, the network at first glance
seems to suffer only minor performance losses.
A detailed quantitative analysis (see below) shows, however,
that the network performance  is diminished by $\approx 30$ \%
due to the detector filter. In view of the severe phase space
limitations caused by the detector cut the network performance
on the filtered sample is nevertheless remarkable.

Test calculations using three-dimensional ($p_x, p_y, p_z$)
information as input rather than the reduced ($p_t, p_z$) set,
did not yield a significant performance improvement.
On the contrary, the use of cartesian momenta
is not feasible: For the proper definition of the $x$ and $y$ directions
the reaction plane has to be determined. Experimentally this is
a complicated task which can only be achieved within an uncertainty of
approximately 30 degrees \cite{gut89}.

The performance of the network can be quantified by computing the
average difference between the reconstructed and the
true impact parameter:
\begin{displaymath}
\Delta b \;=\; \frac{\sum\limits_{i=1}^{N_{\rm events}}
		| b_i^{\rm true} - b_i^{\rm out} |}{N_{\rm events}}
\end{displaymath}
In the first test case ($10 \times 10$ input grid, 10 hidden units)
the network achieved $\Delta~b~=~0.25 $ fm for the unfiltered and
$\Delta~b~=~0.35 $ fm for the filtered input.

How does the performance vary with the dimension of the input grid
and the number of hidden units?
In figure \ref{perform}a) the performance $\Delta b$ is plotted versus
the input grid dimension
(the square root of the number of input components)
for the unfiltered and filtered data with and without
adaption of the input boundaries to the available phase space.
In the region of optimum network performance this adaption has little
effect; only for very small input grid dimensions is a large effect
(a performance gain by 40\%) obtained.

Far more important is the input grid dimension itself.
In the limit of a
($2 \times 2$) input grid one reduces the input basically to the ratio
of transverse to longitudinal  degrees of freedom. This corresponds
to the well known directivity criterion for impact parameter selection.
Increasing the input grid dimension from 2 to 5 improves the performance
by a factor of up to two. This indicates that there are correlations
present in the ($p_z,p_t$) distribution of the event, which are sensitive
to the impact parameter and are {\em not} covered by the directivity
observable or any other related ratio of transverse to longitudinal
degrees of freedom.
However, no significant performance gain is obtained when increasing
the input grid dimension beyond 10.
This finding is important,  since it shows that a further increase
in information presented to the network (via a larger input grid) does
not lead to a better reconstructive power for the impact parameter. Therefore
one can conclude that all ($p_z, p_t$) correlations useful for the
impact parameter reconstruction can be contained in a $5 \times 5$
grid of equidistant bins covering the relevant (detectable)
momentum space area.
The performance loss due to the detector filter
remains constant for dimensions larger than 4 and increases for those
smaller than 4.

In figure \ref{perform}b) the performance $\Delta b$ is again plotted
versus the input grid dimension. Filtered input with asymmetric binning
was used. Two calculations are shown, one with 10 hidden units and one
without any hidden units, where the input components were connected
directly to the output unit (also called simple perceptron).
In the latter case the learning rule becomes particularly simple,
as the input components replace the hidden units.

For input grid dimensions larger than 5,
the number of hidden units seems to be totally irrelevant,
so that the simple perceptron suffices.
In essence the task for the network appears to be a linear one.
This simple network configuration has the great advantage that one
can study its working mechanisms in great detail:
The reconstructed impact parameter is obtained essentially
by computing the scalar
product between the input vector (a one-dimensional representation of
the $10 \times 10$ input grid) and a weight vector.
An ``optimal'' set of weights minimizing the average error for all events
of the training set could in fact be determined via calculation and
inversion of a Hessian matrix. A numerical test confirmed this.
For nonlinear systems with hidden units, however, grossly deteriorated
generalization capabilities would be expected. For better comparability,
the learning process was therefore used in all cases.

\begin{figure}
\centerline{\psfig{figure=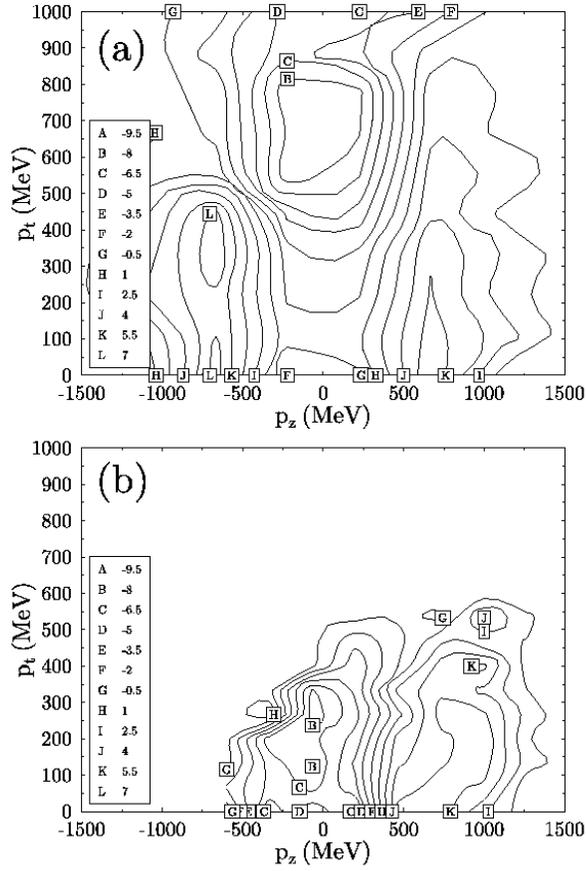,height=15cm}}
\caption{\label{weights} Weight vector mapped onto the representation
of the input grid for unfiltered (top) and filtered (bottom) input.
The weight distribution is directly correlated to the physical
content of the input. }
\end{figure}

In figure \ref{weights} the weight vector of a
($10 \times 10$) network without hidden units
is mapped onto the representation
of the input grid as displayed in figure \ref{pattern}, third row.
Frame a) shows the distribution for unfiltered input, whereas frame
b) shows the distribution for the filtered input.
The distribution of weights correlates directly to the presented
physical input. Areas in the ($p_z,p_t$) plane associated with
small impact parameters are assigned negative weights
(contour lines A-G), whereas those associated
with large impact parameters carry positive weights (contour lines H-L).
This specific
distribution becomes intuitively clear if one remembers that the scalar
product
of input vector and weight vector is simply the sum of the number
of hits in each momentum bin multiplied by their corresponding weights.
Since the result of this sum is the reconstructed impact parameter,
hits in {\em peripheral} areas must contribute large positive values
whereas hits in {\em central} areas must compensate with negative
values.

For the application of the proposed technique on experimental data
the neural network needs to be trained on a data-set with known
impact parameters, i.e. a data-set generated by a transport model.
This model must be able to accurately reproduce the final state
of the heavy ion collision, most importantly the particle and
fragment multiplicities
and the respective double differential cross sections in ($p_t,p_z$) space.
This procedure, however, creates a model bias. It's magnitude can be estimated
by generating two sets of weights from two different
transport models. Then the network can be applied with both sets
on the same experimental data and the difference
in the reconstructed impact parmeters
yields an estimate on the model bias involved.

In summary,
we have applied neural network techniques to the problem of impact
parameter determination in relativistic heavy ion collisions.
As input the transverse and longitudinal momentum distributions
of all outgoing (or detectable) particles were discretized in
two-dimensional grids of varying size. The neural network approach
yields an improvement of a factor of two as compared with classical
techniques. All information necessary to achieve this improvement is
present in a $5\times 5$ grid in ($p_t,p_z$) space. Simple network
algorithms suffice, hidden units are not necessary.

\end{document}